\documentclass{aa}
\usepackage{graphicx}

\begin{document}

\def\spose#1{\hbox to 0pt{#1\hss}}
\def\lta{\mathrel{\spose{\lower 3pt\hbox{$\mathchar"218$}}
     \raise 2.0pt\hbox{$\mathchar"13C$}}}
\def\gta{\mathrel{\spose{\lower 3pt\hbox{$\mathchar"218$}}
     \raise 2.0pt\hbox{$\mathchar"13E$}}}
\def\Msun{{\rm M}_\odot}
\def\msun{{\rm M}_\odot}
\def\Rsun{{\rm R}_\odot}
\def\Lsun{{\rm L}_\odot}
\def\half{{1\over2}}
\def\RL{R_{\rm L}}
\def\zs{\zeta_{s}}
\def\zR{\zeta_{\rm R}}
\def\dJJ{{\dot J\over J}}
\def\dMM{{\dot M_2\over M_2}}
\def\tKH{t_{\rm KH}}
\def\eck#1{\left\lbrack #1 \right\rbrack}
\def\rund#1{\left( #1 \right)}
\def\wave#1{\left\lbrace #1 \right\rbrace}
\def\dd{{\rm d}}
\def\new#1{{#1}}

\title{A Low/Hard State Outburst of XTE J1550-564}

\author{T. Belloni\inst{1}
	\and
	A.P. Colombo\inst{1,2}
	\and
	J. Homan\inst{1}
        \and
        S. Campana\inst{1}
        \and
        M. van der Klis\inst{3}
}

\offprints{T. Belloni}

\institute{INAF -- Osservatorio Astronomico di Brera,
	Via E. Bianchi 46, I-23807 Merate (LC), Italy\\
	\email{belloni@merate.mi.astro.it,
               colombo@merate.mi.astro.it,
               homan@merate.mi.astro.it,
               campana@merate.mi.astro.it}
   \and
        Dipartimento di Scienze, Universit\`a dell'Insubria,
	Via Valleggio 11, I-22100, Como, Italy
   \and
        Astronomical Institute ``A. Pannekoek'' and Center for High-Energy
	Astrophysics, University of Amsterdam, Kruislaan 403, 
	1098 SJ Amsterdam, the Netherlands\\ 
        \email{michiel@astro.uva.nl}
}

\date{Submitted to Astronomy \& Astrophysics, 27 February 2002}

\abstract{We present the results of the analysis of 11 RXTE/PCA observations
	of the January 2002 outburst of the black-hole transient 
	XTE J1550-564. The outburst was rather short, with an e-folding
	time of $\sim$11 days in the PCA band 3-25 keV. 
	The source was seen in an extreme low/hard state. The energy spectra
	could be fitted with a 
        \new{power law slightly steepening ($\Gamma\sim 1.4$ to $\sim$
	1.5) in time, with a high-energy cutoff decreasing from $\sim$300 keV
	to $\sim$100 keV and some additional
	features between 6 and 7 keV, while the 3-150 keV flux
	decreased by a factor of 17.} 
        The power spectra were fitted with two broad Lorentzian
	components, whose characteristic frequencies decreased with
	the source flux, and whose fractional rms contribution was
	rather constant. 
	The timing parameters derived can be identified with those
	observed in other systems.
	These results show that, although during previous, more luminous,
	outbursts, a very complex behavior was observed,
        at low luminosities XTE J1550-564 behaves in a 
	way comparable to most Black-Hole Candidates.

} 

\maketitle

\section{Introduction}

The phenomenology of the X-ray emission from Black Hole Candidates (BHC)
is known to be rather complex, especially that of transient systems. 
Despite the classification into
spectral/timing states that emerged in the last decade (van der Klis 1995;
Tanaka \& Lewin 1995; Belloni 2001), recent observations have revealed
a pattern underlying transitions between states more complex than
previously realized (see e.g. Belloni 1998, 2001; Homan et al. 2001; 
Wijnands \& Miller 2002; Campana et al. 2002).
The large database of observations accumulated with the Rossi X-ray
Timing Explorer (RXTE) greatly contributed to this complexity.
In particular, in the timing domain, complex features are present in 
the Power Density Spectra (PDS) of these systems, both narrow and broad
(see Belloni et al. 2002 and references therein).

The black hole candidate XTE J1550-564  was discovered 
as a bright X-ray transient with the 
All-Sky Monitor (ASM) on board RXTE in September 1998 (Smith 1998). 
Shortly after that the optical and radio counterparts were discovered 
(Orosz et al. 1998; Campbell-Wilson et al. 1998); 
a superluminal ejection has been observed in the radio 
(Hannikainen et al. 2001). 

The mass of the compact object in this binary system is estimated to be 
$\sim$10 M$_\odot$ (Orosz et al. 2002). Its distance has been assumed 
to be 6 kpc (Sobczak et al. 1999),
although there is quite some uncertainty over this value (see e.g. Orosz
et al. 2002).
To date the source was observed to be in outburst four times: 
one major and complex outburst in 1998-99, one in 2000, a small dim
outburst in 2001, and a recent outburst in 2002, also rather dim and
hard (Swank et al. 2002). Low-level activity between the
two latest outburst has also been reported (Swank et al. 2002).

The first outburst was the strongest: in September 1998 the source 
showed a flare that reached 6.8 Crab in the $2-10$ keV band. 
During this outburst, XTE J1550-564 went through all the canonical BHC states.
A complete spectral analysis of these data can be found in Sobczak 
et al. (2000a), where the data are fitted with the ``standard'' model for 
BHC, consisting of the superposition of a disk-blackbody (the soft
component) and a power law (the hard component).
Homan et al. (2001), studying the second part of the outburst in terms of 
color-color diagrams, conclude that to describe the behavior 
of XTE~J1550-564,  a second parameter besides the mass accretion rate
is necessary.
The timing behavior of this outburst was very complex (Cui et al. 1999;
Remillard et al. 1999; Homan et al. 2001).
\new{Radio observations showed that the radio outburst lagged the X-ray outburst
by 1.8 days and reached 375 mJy at 843 MHz (Hannikainen et al. 2001).
An outflow was observed, with an apparent separation velocity $>$2c.
Optical observations are reported by Jain et al. (2001a): 
the correlation between optical and X ray flux levels is poor,
even showing intervals of anti-correlation.}

The 2000 outburst reached a peak flux of $\sim$1 Crab ($1.5-12$ keV):
it also showed state transitions and \new{a timing behavior similar to 
that of the previous outburst}
(Tomsick et al. 2001a; Miller et al. 2001; Kalemci et al. 2001;
\new{J. Miller, priv. comm.}).
\new{Radio observations were reported by Corbel et al. (2001): in the VHS/IS
intervals the radio emission is quenched, while in the Low State
(LS) at the end of the
outburst the radio spectrum is inverted.
Optical and infrared observations showed that the outburst in the V and H 
bands precedes the X-ray outburst by 9-10 days (Jain et al. 2001b).}

Finally, the 2001 outburst showed a $2.5-20$ keV source flux of 
$\sim40$mCrab (measured by RXTE), much weaker than the previous ones.
Spectral and timing properties indicate that the source was in the 
Low/Hard State; energy spectra are described by a power law with a 
photon index of 1.52 and a neutral iron line. 
The rms level of timing noise is about 40\% rms 
(0.01--100 Hz; 2--60 keV; Tomsick et al. 2001b).

As mentioned, in January 2002 
the source became active again, after a long period of low-level activity
(Swank et al. 2002) and it was observed roughly every two
days with RXTE. 
\new{Radio observations were reported by Corbel et al. (2002),
who reported a flat radio spectrum, consistent with a LS.}
As the source was reported to be in the low/hard state,
and since the behavior of BHC at low luminosity is usually very similar
between different sources (see e.g. Belloni et al. 2002),
the analysis of these observations can provide the required link between
this source and other systems. 

\section{Data analysis}

We analyzed a set of 11 public TOO observations of XTE J1550-564 
made with the RossiXTE satellite (Bradt et al. 1993) 
in January 2002. The log of the 
observations is shown in Table 1. RXTE observed the source a few
more times after the last observation presented here, but the 
source had become very weak at those times and we decided to
stop our analysis here. 
We analyzed the data from the Proportional Counter Array (PCA; Jahoda
et al. 1996) instrument \new{and from the High Energy X-ray Timing Experiment
(HEXTE: Rothschild et al. 1998).}
For each observation we produced a 
power density spectrum, X-ray colors and an energy spectrum, following
the procedures outlined below.

\begin{table}
   \caption[]{Observation log. PCA rates are renormalized to one PCU
        For the definition of X-ray colors, see text.
        }
       \label{tab1}
   $$
   \begin{array}{lcccccc}
     \hline
     \noalign{\smallskip}

{\rm Obs.}   &{\rm Date}   &{\rm St. Time} &{\rm Exp.} &{\rm PCA Rate} & {\rm HR}_1 & {\rm HR}_2\\
       &{\rm (2002)} & {\rm (UT)}     &{\rm (s)}  & {\rm c/s}  &      &     \\
     \noalign{\smallskip}
     \hline
     \noalign{\smallskip}
{\rm A}      & {\rm Jan}\, 10& 21{:}19    & 4096& 202.4& 1.27 & 0.15    \\
{\rm B}      & {\rm Jan}\, 13& 02{:}34    & 3584& 175.2& 1.28 & 0.15    \\
{\rm C}      & {\rm Jan}\, 15& 02{:}07    & 4096& 145.7& 1.28 & 0.15    \\
{\rm D}      & {\rm Jan}\, 17& 09{:}16    & 2048& 118.2& 1.27 & 0.15    \\
{\rm E}      & {\rm Jan}\, 19& 10{:}50    & 2560&  96.1& 1.28 & 0.15    \\
{\rm F}      & {\rm Jan}\, 21& 18{:}18    & 2048&  77.5& 1.30 & 0.15    \\
{\rm G}      & {\rm Jan}\, 23& 05{:}16    & 4096&  59.2& 1.27 & 0.14    \\
{\rm H}      & {\rm Jan}\, 25& 20{:}53    & 2048&  48.8& 1.29 & 0.14    \\
{\rm I}      & {\rm Jan}\, 26& 23{:}38    & 2048&  41.6& 1.30 & 0.14    \\
{\rm J}      & {\rm Jan}\, 28& 20{:}03    & 2560&  32.7& 1.27 & 0.13    \\
{\rm K}      & {\rm Jan}\, 31& 19{:}14    & 3072&  22.0& 1.17 & 0.11    \\
     \noalign{\smallskip}
     \hline
\end{array}
   $$
\end{table}
\subsection{All-sky monitor}

The All-Sky Monitor (Levine et al. 1996) on board RXTE 
reaches a sensitivity of $\sim 5-10$ mCrab (1.3-12 keV)
over one day. Fig. 1 shows the light curve of XTE~J1550--564 across the 2002 
outburst\footnote{http://xte.mit.edu/lcextrct/}, with a bin size of one day.
The outburst likely started during Dec. 2001, but the closeness of the Sun 
made monitoring of the source difficult due to scattered solar
X--rays. The situation improved in Jan. 2002, when a clear decrease in the 
source flux was observed testifying to the latest stages of an 
outburst. A linear or exponential decay can fit the data equally 
well, with a count rate decrease of 0.3 per day or an e-folding 
time of $\sim 15$ d.

\begin{figure}
\centering
\includegraphics[width=0.5\textwidth]{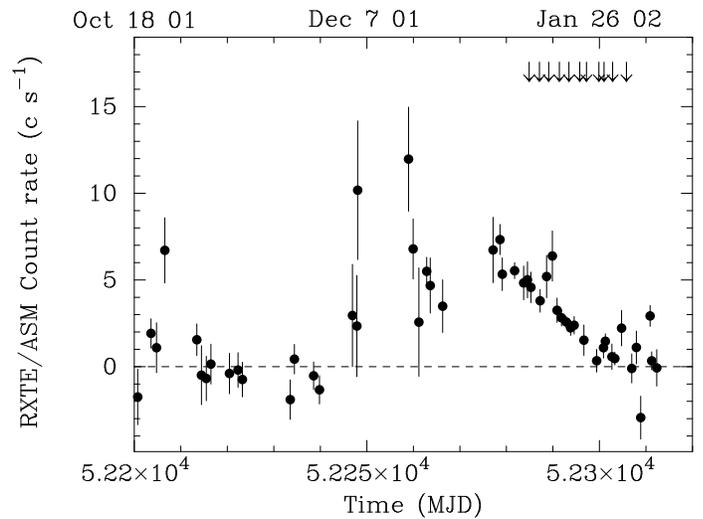}
  \caption{RXTE/ASM light curve of the 2002 outburst of XTE J1550-564.
		Each point represents the average of the scans within one day.
		The times of the pointed observations analyzed here are shown
		as vertical arrows.
	  }
  \label{asm}
\end{figure}

\subsection{PCA/HEXTE: spectral analysis}

To assure that the PCA energy spectra for each individual observation would
not contain data from different count rate levels and to identify possible
problems with individual PCUs, 
we produced PCA light curves with a 1-s time resolution using {\tt Standard1}
data. All light curves appear to be rather stable in average count rate,
with the clear presence of strong noise in the form of rapid flaring.
The total count rates of the 11 observations 
are shown in Table 1, together with two X-ray colors
defined as $HR1$=B/A $HR2$=C/A, where A, B and C are the net counts in 
the channel ranges 0-14 (2.0-6.1 keV), 15-37 (6.1-15.8 keV) 
and 38-46 (15.8-19.4 keV) respectively. The count rates and 
colors were extracted from the
{\tt Standard2} data, after estimation of the background (see below).
These bands are the same used by Homan et al. (2001).
The total PCA light curve is shown in Fig.~2: 
an exponential 
fit gives an e-folding time of 10.6$\pm$0.8 days, a figure lower than
that estimated from the ASM light curve, possibly due to the difference
in energy range.
Both X-ray colors appear to be very stable throughout the observing period,
despite the count rate decrease by a factor of ten (in the 3.0-25.0 keV
band), indicating that the
overall spectral shape does not vary much. However, the last spectra tend to be
slightly softer. This is confirmed by the results of the spectral fits.

We extracted PCA energy spectra from {\tt Standard2} data, which have
an intrinsic time resolution of 16 seconds.
We accumulated a single energy spectrum for each observation, selecting
all available PCUs with the exception of PCU0, to avoid problems due to the
loss of the propane layer in that unit. The spectra were extracted using
standard procedures with FTOOLS V5.1. The background was estimated using
{\tt pcabackest} V2.1e and subtracted from the data. PCA response matrices
were produced with {\tt pcarsp} V7.11. For all spectra, only channels
4-52, corresponding to 3-25 keV were used.
\new{We extracted HEXTE energy spectra from both instrument clusters using
standard FTOOLS V5.1 procedures. For the fits, we considered only channels
10-50, corresponding to 18-150 keV.}
For the fits, a systematic error of 0.75\% was added to each channel.
We fitted the energy spectra in the range 3-150 keV 
with a model consisting of a power law \new{with a high-energy cutoff},
a smeared edge and a gaussian emission line. The interstellar absorption 
was fixed at 8.5$\times 10^{21}$cm$^{-2}$ (Tomsick et al. 2001a).
\new{For HEXTE, we used only cluster B, besides for observation C for which
the cluster B background count rate was abnormally high and prevented the
production of a meaningful net spectrum. Results using cluster A are 
compatible with those reported here.} 
Note that our lower boundary of 3 keV would make it difficult to detect
the presence of a very soft component.
\new{For observations A through G the Gaussian line was
not required by the fit, while for observations H through K the edge was
not needed.}
As the energy of line and edge proved to be very stable through all 
observations \new{where they were required}, 
they were fixed to 6.4 keV and 7.0 keV respectively. Note
that these features can be partly the result of and/or be influenced 
by instrumental effects. 
The best fit values for the power-law photon index $\Gamma$, 
\new{the cutoff energy E$_c$,} 
and the integrated flux, plus the reduced $\chi^2$, are shown in Table 2.
Some of the fits are formally unacceptable. \new{Inspection of the residuals
shows that the problems are mostly due to a rather large scatter of the
HEXTE points, with the exception of 
observation K, for which a large-scale modulation in the residuals indicate
that a more complex model is probably needed.}
\new{As one can see from Table 2, there is some evidence of a steepening of the
power-law index (from $\sim$1.4 to $\sim$1.5) and of a decrease in 
cutoff energy (from $\sim$300 keV to $\sim$100 keV) while the observed
3-150 keV flux decreases by more than one order of magnitude.}
\begin{table*}
   \caption[]{Best-fit spectral and timing parameters for the observations 
        analyzed here.
        Columns are: observation ID; spectral parameters: 
        power-law photon index, 
         \new{cutoff energy E$_{\rm c}$ in keV,}
        3-150 keV \new{unabsorbed} flux in 10$^{-9}$erg 
        cm$^{-2}$ s$^{-1}$, reduced $\chi^2$; timing parameters:
        centroid frequency and HWHM of the
        first Lorentzian, centroid frequency and HWHM of the second
        Lorentzian, reduced $\chi^2$ with number of degrees of freedom,
	total integrated fractional rms (0.002-10.0 Hz).
	All errors and upper limits are at 90\% confidence.
        }
       \label{tab2}
   $$
   \begin{array}{lccccccccccccc}
     \hline
     \noalign{\smallskip}

{\rm Obs.}&\Gamma     &E_c&{\rm Flux}&\chi_r^2&\nu_1^c       & \Delta_1     & \nu_2^c   & \Delta_2     &\chi_r^2 &{\rm \% rms}\\
          &           &{\rm (keV)}&    &{\rm (44 dof)}&{\rm (Hz)}          &  {\rm (Hz)}        &  {\rm (Hz)}   &  {\rm (Hz)}        &  {\rm (dof)} & \\
     \noalign{\smallskip}
     \hline
     \noalign{\smallskip}
{\rm A}   &1.39\pm 0.01&339\pm 52       &12.15&0.88&0.016\pm 0.006&0.087\pm 0.011&<0.024   &2.132\pm 0.075& 0.99  (181) &37\\
{\rm B}   &1.42\pm 0.01&>566            &10.56&1.29&0.012\pm 0.008&0.088\pm 0.010&<0.023   &1.957\pm 0.075& 1.09  (181) &36\\
{\rm C}   &1.41\pm 0.01&>411            &8.65 &1.33&0.016\pm 0.008&0.080\pm 0.010&<0.022   &1.863\pm 0.074& 1.18  (181) &35\\
{\rm D}   &1.42\pm 0.02&>474            &6.60 &1.23&< 0.020       &0.075\pm 0.015&<0.047   &1.655\pm 0.131& 0.96  (181) &35\\
{\rm E}   &1.39\pm 0.02&289^{+139}_{-78}&5.64 &0.70&< 0.015       &0.062\pm 0.012&<0.026   &1.579\pm 0.099& 0.82  (181) &36\\
{\rm F}   &1.41\pm 0.02&194^{+75}_{-50} &3.80 &1.19&< 0.011       &0.048\pm 0.009&<0.042   &1.448\pm 0.108& 0.94  (181) &34\\
{\rm G}   &1.43\pm 0.02&354^{+200}_{-126}&3.61 &0.77&< 0.006       &0.048\pm 0.005&<0.015   &1.472\pm 0.076& 1.22  (181) &36\\
{\rm H}   &1.50\pm 0.02&>178            &2.21 &1.21&< 0.030       &0.064\pm 0.025&<0.030   &1.318\pm 0.209& 1.05  (181) &35\\
{\rm I}   &1.45\pm 0.03&147^{+114}_{-44}&1.81 &1.05&< 0.019       &0.026\pm 0.008&<0.030   &0.992\pm 0.111& 1.16  (145) &35\\
{\rm J}   &1.49\pm 0.03&129^{+229}_{-51}&1.29 &1.15&< 0.011       &0.027\pm 0.009&<0.044   &0.907\pm 0.113& 1.03  (145) &36\\
{\rm K}   &1.49\pm 0.05&64^{+50}_{-19}  &0.710 &1.37&< 0.014       &0.020\pm 0.006&<0.025   &0.630\pm 0.073& 1.14  (145) &35\\
     \noalign{\smallskip}
     \hline
\end{array}
   $$
\end{table*}

\subsection{PCA: timing analysis}

The PCA light curves show strong aperiodic variability, characteristic
of the low/hard state.
For each observation, we produced a Power Density Spectrum (PDS) in
the following way. Individual Leahy-normalized (Leahy et al. 1983)
PDS were produced from 512-second 
stretches of data with a time resolution of 512 s$^{-1}$, selecting
PCA channels 0-35, corresponding to roughly 2-15 keV. The analysis
of data above 15 keV did not show significant differences, both
in frequencies and rms amplitudes. These
PDS were averaged together and the contribution from the Poissonian
statistics was subtracted (Zhang et al. 1995). Then the average PDS
was normalized to squared fractional rms (Belloni \& Hasinger 1990a).
Since the signal rapidly vanishes at high frequency, we excluded
from the analysis bins above 40 Hz for observations A-H, and above
10 Hz for observations I-K. The resulting PDS can be seen in 
Fig.~3, plotted in a $\nu$P$_\nu$ representation.

Each PDS was then fitted with a model consisting of Lorentzian
components (see Belloni et al. 2002). Fits were
performed with XSPEC V.11.1.0. For all power spectra,
two Lorentzians were sufficient to achieve an acceptable fit. The best
fits are shown in Fig. 3, and the best fit frequency parameters 
(centroid frequencies $\nu_{1,2}^c$ and HWHM $\Delta_{1,2}$) 
are reported in Table 1, together with 
the corresponding reduced $\chi^2$. The best fit is almost always
reached for two zero-centered Lorentzians: nevertheless we did not
want to force the centroid frequency of the Lorentzians to zero (see
Belloni et al. 2002). From Fig.~2, 
it is evident that as the outburst proceeds and the source count rate 
decreases, the peak of both Lorentzians in $\nu$P$_\nu$ moves
to lower frequencies. In order to quantify this, we computed the
characteristic
frequencies $\nu_{1,2}$ of the peaks in $\nu$P$_\nu$ according 
to Belloni et al. (2002) and plotted the values as a function of time
during the outburst (see Fig.~2). Notice from Table 2 
that the total fractional rms
integrated in the 0.002-10.0 Hz is remarkably constant, as it is also
apparent from the evolution of the power spectra in Fig. 3.
\new{Our fits show that also the rms of the single Lorentzian components 
remains constant throughout the observed part of the outburst.}

\begin{figure}
\centering
\includegraphics[width=0.5\textwidth]{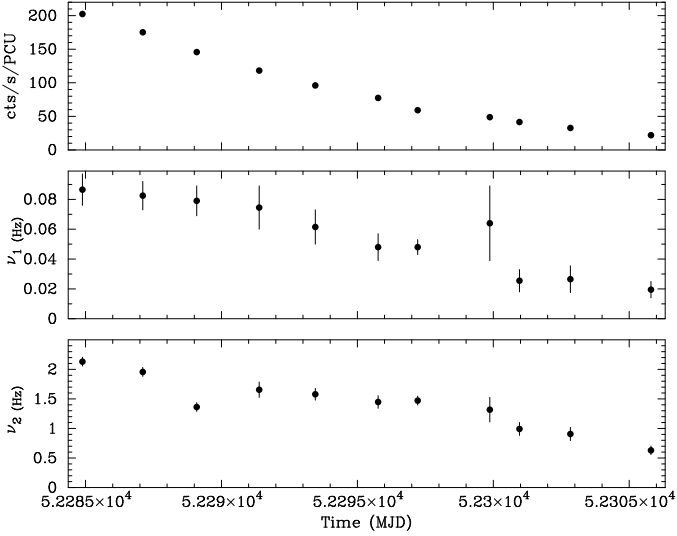}
  \caption{Top panel: overall PCA light curve (3.0-25 keV) corresponding to the 
           observations analyzed here. Middle and low panel: timing
	   evolution of the characteristic frequencies of the two
	   timing components in the PDS of the source.
	  }
  \label{licu}
\end{figure}
\begin{figure}
\centering
\includegraphics[width=0.5\textwidth]{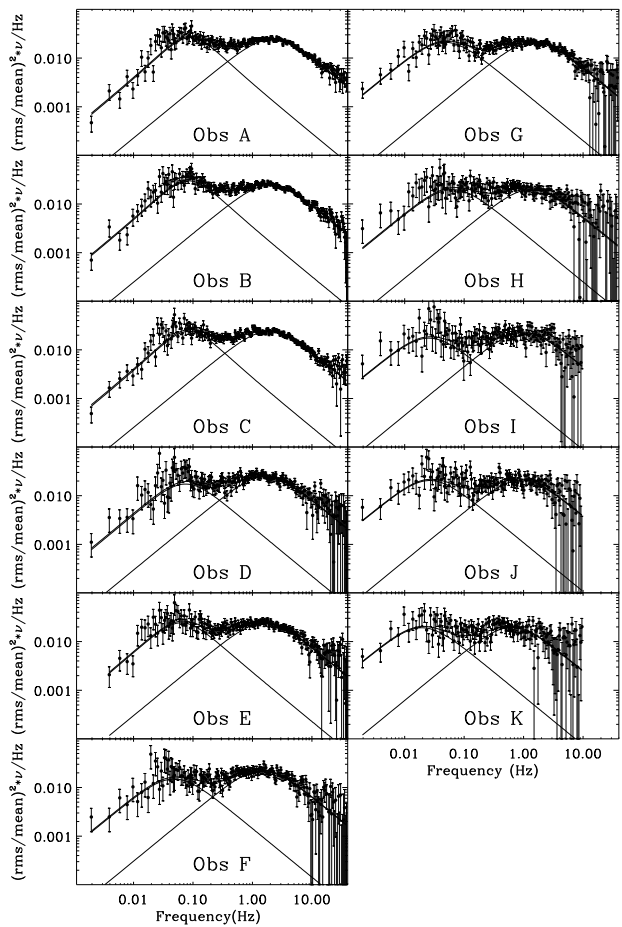}
  \caption{Power Density Spectra for all the observations presented here
           in $\nu$P$_\nu$ form. The best fit models are also shown.           
	  }
  \label{pds}
\end{figure}
\section{Discussion}

The first and major outburst of XTE J1550-564 in 1998-1999 has provided
important information about the connection between spectral/timing states
in black-hole transients indicating that a second
parameter besides the mass accretion rate is governing state transitions
(Homan et al. 2001). This outburst has shown all major source states,
in addition to an extremely complex phenomenology in the aperiodic time
variability, including strong 1-10 Hz QPO peaks with multiple harmonics
and high-frequency QPOs between 100 and 280 Hz (Remillard et al. 1999;
Homan et al. 2001). The following outburst in 2000 also showed some
state transitions (Tomsick et al.
2001a; Kalemci et al. 2001; Miller et al. 2001). At the end of both
outbursts, however, the source entered a low/hard state, where the energy
spectrum is dominated by the hard power-law component and the power
density spectrum shows a strong band-limited noise component.
During the short weak outburst of 2001, a PCA observation showed a
low/hard state, with a hard energy spectrum (power law with $\Gamma$=1.52)
and a strong (40\% fractional rms) band-limited noise (Tomsick et 
al. 2001b). Violent flaring activity was also reported.

The outburst analyzed here is quite different from the first two and
similar to the 2001 one. Throughout all the observations the source was
clearly in the low/hard state, both from the spectral and timing analysis.
\new{The unabsorbed 3-150 keV flux decreased from 
1.2$\times 10^{-8}\,$erg cm$^{-2}$s$^{-1}$ to 
7.1$\times 10^{-10}\,$erg cm$^{-2}$s$^{-1}$. At a distance of 6 kpc (Sobczak et
al. 1999) these correspond to 5.0-0.2 $\times 10^{37}\,$erg s$^{-1}$.
For comparison with Tomsick et al. (2001a), the corresponding
3-25 keV absorbed fluxes are  2.5$\times 10^{-9}\,$erg cm$^{-2}$s$^{-1}$ and 
3.5$\times 10^{-10}\,$erg cm$^{-2}$s$^{-1}$.
The PCA LS observations from 2000 covered roughly the same range
(Tomsick et al. 2001a), and the
observation of the 2001 outburst saw the source at a flux intermediate
between these limits (Tomsick et al. 2001b).
}
Both the timing (characteristic frequencies) and spectral parameters indicate
that indeed we are observing a rather low-flux hard state, where the
energy spectrum is very hard and the variability is shifted to low frequencies.
\new{Notice that our power-law component is flatter than observed during the
whole 2000 outburst, even at similarly low flux levels.
As the source flux (and count rate) decreases, we see the low-energy part
of the spectrum steepen slightly, and the cutoff moves to lower energies.
With the caveat of rather large error bars, a large change in cutoff
energy seems to be observed: in the framework of a Comptonization model,
this would imply that, in addition to the electron temperature (measured
through the observed cutoff), also the optical depth of the Comptonizing cloud 
increased.
Moreover, although we do see a relation between the characteristic frequencies
and spectral properties, they seem to be less strong than those observed
at higher luminosities (see also below).
}
\new{Notice that we are probably not covering the whole outburst (see Fig. 1),
and we cannot exclude the presence of different source states before our
observations took place. Indeed, Tomsick et al. (2001a) 
report that in the
2000 outburst the transition to the LS took place at a slightly higher 
3-25 keV luminosity
that what we observe in our first pointing.}

The PDS can all be fitted with a very simple model consisting of only two
Lorentzian components, mostly with zero centroid frequency. This is consistent
with recent applications of the Lorentzian models to BHC in the hard
state (Belloni et al. 2002; Pottschmidt et al. 2002). However,
since only two components are formally needed here, the question arises
of the identification of these components in the scheme described by
Belloni et al. (2002). As van Straaten et al. (2002) 
pointed out, this is only possible by comparing a number of observations.
Indeed, we can interpret our low-frequency component as $L_b$ and our
high-frequency component as $L_\ell$ (see Belloni et al.
2002). Fig. 4 shows our $\nu_{\rm 1}$ and $\nu_{\rm 2}$ values in 
a $\nu_\ell$ vs. $\nu_{\rm b}$ plot like the 
one in van Straaten et al. (2002). The points for GX 339-4 from Nowak et al.
(2002) and those for XTE J1118+480 from Belloni et al.
(2002) are also plotted. Our points agree rather well
with those from XTE J1118+480 and with an extrapolation of the 
highest GX 339-4, but are a factor of two below the
the lowest GX 339-4 points.
Identifying the low-frequency component with $L_{\rm b}$ allows us to check
the expected correlation between flat-top level (defined here as the value
of the $L_{\rm 1}$ component PDS at our lower frequency boundary of 
0.002 Hz) and $\nu_{\rm b}$ (Belloni
\& Hasinger 1990b; Belloni et al. 2002). Fig. 4,
where the points from Belloni et al. (2002)
are also plotted, shows that the expected correlation is followed with the
right $-$1 slope.
 
Vignarca et al. (2002) find that in a number of BHC
there is a correlation between power-law slope $\Gamma$ and frequency of 
the low-frequency QPO. For XTE J1550-564, for QPO frequencies below 2 Hz this
correlation is positive and flattens at low QPO frequencies (see also
Sobczak et al. 2000b). This means that at low flux, when $\Gamma$ is as 
low as 1.5, one expects that large variations in the QPO frequency 
correspond to small corresponding variations in $\Gamma$.
\new{Our values correspond to a low-frequency extension of the points from 
Sobczak et al. (2000b); however, they show a weak anticorrelation, opposite
to what we expected.}

In conclusion, during its short outburst in January 2002, XTE~J1550-564
showed spectral and timing features typical of an extreme low/hard state,
\new{at least in the PCA/HEXTE observed part of the outburst}.
Its characteristic parameters in the timing and spectral domains are
compatible with those of many other BHC. This indicates that this source,
at least at low flux values (and possibly also low values of the mass
accretion rate) behaves in the expected way,
\new{although the weak spectral variations we observe are not in the
direction of what is usually seen from these
systems (see also Tomsick et al. 2001a)}. 
This allows to link the properties of this source with those of
other systems and indicates that bright BHCs while at low luminosities
behave in a rather coherent way.

\begin{figure}
\centering
\includegraphics[width=0.5\textwidth]{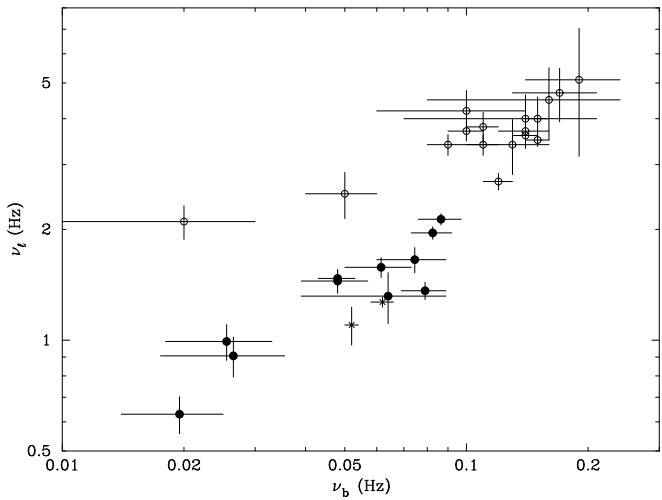}
  \caption{Correlation between the characteristic frequencies of the
	two Lorentzian components fitted to the PDS of XTE J1550-564 (filled
	circles). For comparison, the $\nu_b$ and $\nu_\ell$
        frequencies for
	XTE J1118+480 (crosses: Belloni et al. 2002) and 
	GX 339-4 (open circles: Nowak et al. 2002) are shown.
	  }
  \label{nunu}
\end{figure}
\begin{figure}
\centering
\includegraphics[width=0.5\textwidth]{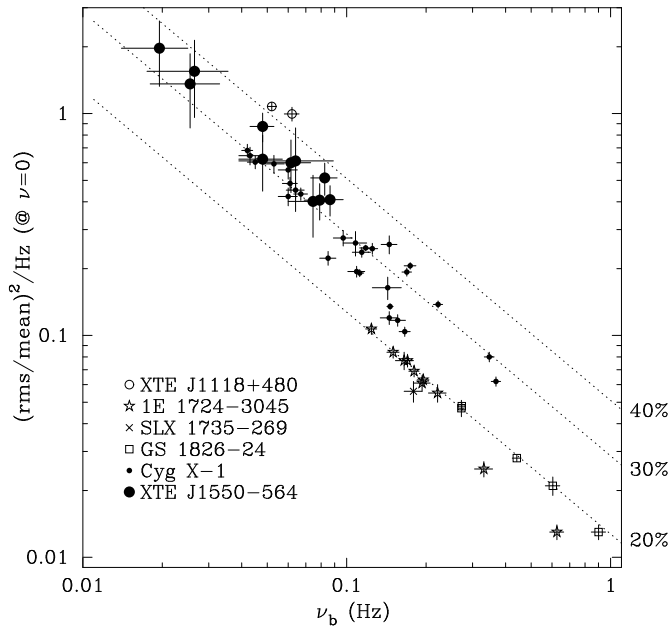}
  \caption{Correlation between $\nu_b$ and flat-top level (see text). 
           The XTE J1550-564
	   points are marked with large filled circles. All other symbols
	   are the same as Fig. 10 of Belloni et al. 
	   (2002). The dotted lines represent constant total fractional
	    rms. Numbers on the $y$ axis are squared rms times two 
	   (see Belloni et al.2002).
	  }
  \label{bhall}
\end{figure}

\begin{acknowledgements}
This work was done with support from the European Commission, the Training
and Mobility of Researchers (TMR) research network programme.
T.B. thanks the Cariplo Foundation for financial support. J.H. 
acknowledges support from Cofin-2000 grant MM02C71842. S.C. thanks 
CNR and ASI for support. M.K. acknowledges
support from the Netherlands Organization for Scientific Research (NWO).
\end{acknowledgements}


\end{document}